# A giant overshoot effect in the Janssen granular column

G. Ovarlez, C. Fond(*), E. Clément
*Laboratoire des Milieux Désordonnés et Hétérogènes - UMR 7603 - Université Pierre et Marie Curie - Boîte 86 - 4, Place Jussieu,F-75252 Paris, France*
*(*) I.U.T. Robert Schuman - Département Génie Civil - 72, route du Rhin, B.P. 315, 67411 Illkirch-Graffenstaden, France*

We present new experimental results on the mechanical behavior of static granular assemblies confined in a vertical column. Our measurements confirm, for the first time, the universal Janssen's scaling for the stress saturation curve. We show consequently, in the context of isotropic elasticity, a relation between the Poisson ratio and granular packing fraction. Moreover, using a systematic study of the overshoot effect created by a top mass equal to the saturation mass, we show behaviors reproduced qualitatively by isotropic elastic materials but in the case of a granular assembly of a spectacular amplitude. These experimental results are strong tests for any theory of granular matter.

PACS numbers: 45.05.+x,45.70.Cc,46.25.-y

The mechanical status of granular matter is presently one of the most open and debated question [1]. This state of matter exhibits many unusual mechanical and rheological properties such as stress induced organization at the microscopic [2] or at the mesoscopic [3] level. This issue sets fundamental questions relevant to the understanding of many other systems exhibiting jamming such as dense colloids or more generally soft glassy materials [4, 5]. For practical applications, the quasistatic rheology of granular assemblies is described using a phenomenological approach, based on an elasto-plastic modelling of stress-strain relations [6]. So far, there is no consensus on how to express correctly the macroscopic constitutive relations solely out of microscopic considerations and under various boundary conditions or loading histories. This very basic interrogation was illustrated in a recent debate on how to understand the stress distribution below a sand pile and especially how to account for the dependence on preparation protocols [7]. A new mechanical approach was proposed based on the concept of "fragile matter" [5] and force chains propagation modelling [8]. But recent experiments have dismissed this approach and evidenced results more consistent with the traditional framework of general elasticity [9].

In this letter we obtain new and non-intuitive results on stresses measurements at the bottom of a granular assembly confined in a rigid cylinder. Two kinds of experiments were performed. First, the mass at the bottom of the column is measured as a function of the granular material filling mass. Second, similar measurements are produced with an overweight on the top of the granular material. Similar experiments were previously made by Vanel et al. [10] who showed that contrarily to previous experimental reports, one could obtain quite reproducible data provided a good control of the packing fraction homogeneity and a mobilization of all the friction forces at the walls in the upwards direction. The existence of an overshoot effect was clearly demonstrated then, but no interpretation nor systematic measurement were undertaken. Note that this overshoot effect is also an outcome of the "OSL theory" [8] based on a vault propagation modelling but an important feature of the theory was not evidenced experimentally [10]: the existence of stress oscillations. In the present report, we use the same experimental set-up as in [10] but with a noticeable change in the measurement procedure which we claim is a decisive improvement for the data interpretation.

The grains are dry, noncohesive, and slightly polydisperse (10%) glass beads of diameter $d = 1.5$ mm piled into a vertical steel cylinder. A brass overweight can be added on top of the granular material. Three different column radii are used: $R = 19$ mm, 28 mm and 40 mm. The static coefficient of friction $\mu_s$ between the glass beads and the steel walls is measured using the sliding angle of a three-bead tripod. We found: $\mu_s = 0.25 \pm 0.02$. Two other steel cylinders of radius $R = 19$ mm are used: a rough one ($\mu_s = 0.285 \pm 0.02$), and a smoother one ($\mu_s = 0.22 \pm 0.02$). The column is closed at the bottom by a movable piston (the diameter mismatch is 0.5 mm). A force probe of stiffness $k = 40000$ N.m$^{-1}$ is located under the piston and can be moved at constant driving velocity $V$ via a stepping motor (see Fig. 1 inset).

The experiments were performed on three different pilings: a dense one ($64.5 \pm 0.5\%$) obtained by rain-filling, a loose one ($59.0 \pm 0.5\%$) obtained by using an inner cylinder which is removed slowly after filling, and a medium one ($62.0 \pm 0.5\%$) obtained by vibrating a loose packing.

For *each data point*, the same procedure is repeated: first the column is filled with a mass of beads $M_{fill}$ and prepared according to the desired average packing fraction $\nu$. Then, the vertical translational stage is moved down slowly by a step motor (resolution of 100 nm/step) and the mass on the gauge is monitored. From the mass versus vertical descent data, the apparent mass $M_a$ at the bottom of the column is extracted. We use a very slow descent of the force probe at a velocity $V_0 = 1.5$ $\mu$m.s$^{-1}$, more than ten times smaller than the one used by Vanel et al. [10]. Then, a quasi-static situation is attained were all dynamical oscillations are suppressed. In the first stage of the descent, the granular material does not move (see Fig. 1 inset). The stress probe yields a value : $m(t) = m_0 - (k/g)V_0 t$, where $m_0$ is the apparent mass at time $t = 0$ just after the end of the filling procedure (note that the value of $m_0$ is highly irreproducible from one filling to the other). When the friction at the walls reaches the Coulomb threshold, the material starts to slip and then gently stabilizes at the value $V(t) \simeq V_0$. Then the apparent mass may decrease for a piston descent of another 50 $\mu m$ or more (the precise value depends on the column size) before it reaches a stationary state. This later evolution could be due to mechanical restructuring of the piling caused by the slippage at the boundaries. This evolution to a plateau limit should not be mixed with a dilatancy effect, which would occur much later, for displacements of about a grain size and also would lead to an increase of the apparent mass [10]. We choose to make the apparent mass $M_a$ measurements, just at the end of the linear relaxation regime (see arrow on Fig. 1 inset). Therefore, this new procedure should ensure (i) that the static piling is at the threshold of slippage everywhere at the walls and (ii) that we measure the mechanical properties of the initial piling in the absence of mechanical restructuring. These are major differences with the original procedure of Vanel et al. [10] who extracted the apparent mass at rest, after relaxation from the plateau value. In cylindrical coordinates with origin at the top surface and the cylinder axis being the $z$ axis, the relation, at the slipping onset, between the shear stress $\sigma_{rz}$ and the horizontal stress $\sigma_{rr}$ at the walls is $\sigma_{rz}(r{=}R, z) = \mu_s \sigma_{rr}(r{=}R, z)$ ($\mu_s$ is the Coulomb static friction coefficient between the grains and the wall).

The typical results obtained when the filling mass $M_{fill}$ is varied are shown on Fig. 1. The apparent mass $M_a$ saturates exponentially with $M_{fill}$. When an overweight equal to the saturation mass $M_{sat}$ is added on top of the granular material, $M_a$ increases with $M_{fill}$, up to a maximum $M_{max}$, then decreases slowly towards the saturation mass $M_{sat}$. We also verified that if we add and remove slowly the overweight, we can recover the apparent mass corresponding to the original Janssen's saturation curve within the experimental uncertainties.

The simple model which captures the physics of this saturation phenomenon was provided in 1895 by H.A. Janssen [11]. This model is based on the equilibrium of a granular slice taken at the onset of sliding and predicts a relation between the filling mass $M_{fill}$ and the apparent mass at the bottom $M_a$ of the form:

$$M_a = M_{sat}(1 - \exp(-\frac{M_{fill}}{M_{sat}})) \text{ with } M_{sat} = \frac{\rho \pi R^3}{2K\mu_s} \quad (1)$$

where $\rho$ is the mass density of the granular material, and $K$ is the Janssen parameter rendering the average horizontal redirection of vertical stresses. From a mechanical point of view, a major simplification of this model comes from the assumption that the redirection parameter $K$ would stay constant along the vertical direction. But on the other hand, it provides a clear and simple physical explanation for the existence of an effective screening length $\lambda = R/2K\mu_s$ above which the mass weighted at the bottom saturates. Many saturation profiles were measured for various packing fractions, columns sizes and friction coefficients. Janssen's model may look very elementary but when the apparent mass rescaled by the saturation mass is plotted as a function of the filling mass also rescaled by the saturation mass, we obtain a *universal rescaling* of all data on a curve which is precisely the one predicted by Janssen : $M_a/M_{sat} = f(M_{fill}/M_{sat})$, with $f(x) = 1 - \exp(-x)$ (see Fig. 2). We recall that each data point corresponds to an independent experiment. The results is a priori so surprising that we will try next to understand it in reference to a simple elastic modelling of the granular column. The differences in $M_{sat}$ measured for different densities means that the effective Janssen's coefficient $K$ increases with the average packing fraction (Fig. 2 inset). In other words the efficiency of vertical stresses redirection is better for higher packing fractions and it is a very sensitive result ($\Delta K/K \simeq 5\Delta\nu/\nu$). We also obtain a Janssen's coefficient $K$ independent of the radius $R$ and of the friction coefficient $\mu_s$ (Fig. 2 inset). These results are to the best of our knowledge the *first* experimental validation of this 1895 model.

For experiments with an overweight on the top, we also rescale the data for different column sizes (Fig. 3a) and also find a universal curve $g(x)$ such that $M_a/M_{sat} = g(M_{fill}/M_{sat})$ which means that the data scale with $R^3$, as $M_{sat}$ does. Note that this overshoot goes beyond the possibilities of Janssen's model which would predict a flat curve: $M_a/M_{sat} = 1$. But, as we change density or friction, we find no rescaling of the data. On Fig. 3b inset we display the maximum overshoot value $M_{max}$ rescaled by the saturation mass $M_{sat}$ and show that its value increases with packing fraction and friction at the walls. Thus we can reformulate this result in a very surprising and non-intuitive way: *the more the weight of the grains is screened by the walls, the less the weight of the overload is screened*! Note that a $R^3$ scaling for the apparent masses (valid with and without overweight), rules out theoretical predictions [12] based on the q-model [13]. Furthermore, no stress oscillation is present contrary to the OSL theoretical predictions [10].

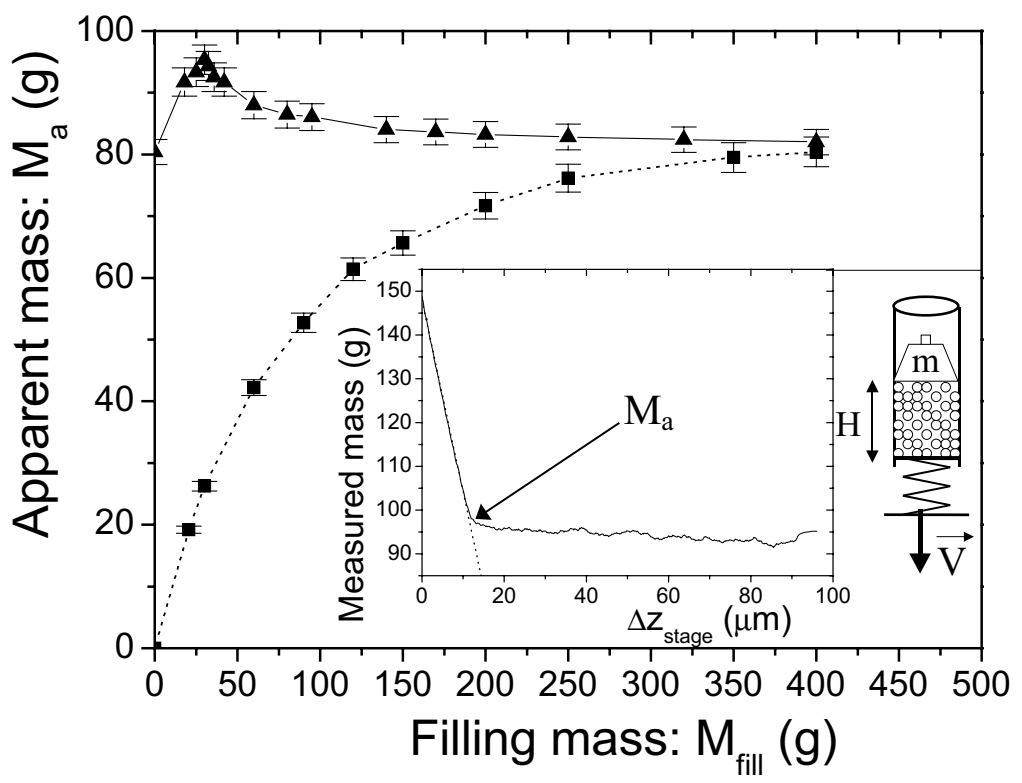

FIG. 1: Apparent mass $M_a$ vs. filling mass $M_{fill}$ for a loose packing in the medium-rough 38 mm diameter column without an overweight (squares), and with a 80.5 g overweight (triangles). Inset: measured mass vs. translation stage displacement $\Delta z_{stage}$, for a loose packing in the smooth 38 mm diameter column; the arrow indicates the measure point; the dotted line corresponds to the linear spring relaxation (slope -k/g). Sketch of the experimental display.

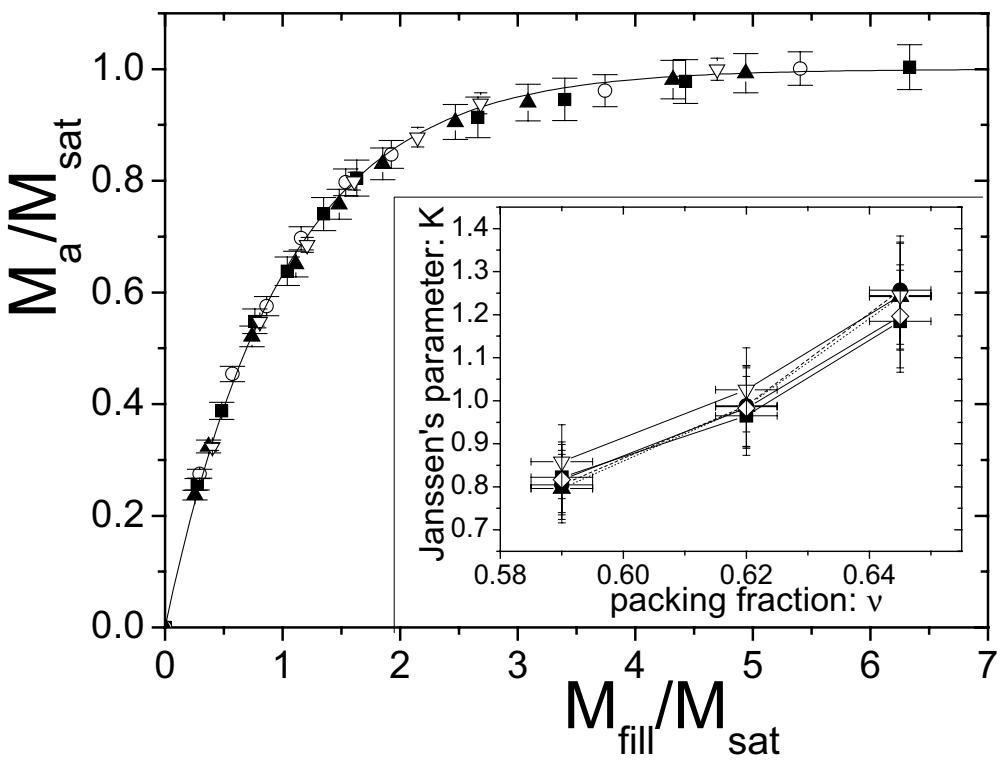

FIG. 2: Apparent mass $M_a$ vs. filling mass $M_{fill}$, rescaled by the saturation mass $M_{sat}$, for a loose packing in the smooth 38 mm diameter column (triangles), for a dense packing in the rough (open circles) and smooth (squares) 38 mm diameter columns, and for a dense packing in the 80 mm diameter column (open down triangles). The results are compared to Janssen's model prediction (line). Inset: Janssen's parameter K extracted from experiments made in medium-rough columns of 3 diameters: 38 mm (squares), 56 mm (circles), 80 mm (triangles), and in a smooth (open down triangles) and a rough (open diamonds) 38 mm diameter column, vs. packing fraction.

We now study in detail the predictions of isotropic homogeneous elasticity. We consider an elastic medium, of Young modulus $E$ and Poisson coefficient $\nu_p$, in a column of radius $R$. Stresses and displacements can be calculated in the limit of high depths $z$ under the assumption that they then should be independent of $z$. The boundary conditions we impose are the Coulomb condition everywhere at the walls ($\sigma_{rz}(r{=}R) = \mu_s \sigma_{rr}(r{=}R)$), and infinitely rigid walls i.e. $u_r(r{=}R) = 0$. Using the stress-strain relation $E\epsilon_{ij} = (1+\nu_p)\sigma_{ij} - \nu_p \delta_{ij}\sigma_{kk}$ and internal equilibrium relation $\partial_i \sigma_{ij} = -\rho g_j$ we find $\sigma_{rr} = \sigma_{\theta\theta} = (\nu_p/(1-\nu_p))\sigma_{zz}$ with $\sigma_{zz}^{sat}(r,z) = -\frac{\rho g R}{2\mu_s}$. The asymptotic displacements are $u_z(r,z) = -\frac{1+\nu_p}{2E}\rho g r^2 - \frac{1-\nu_p-2\nu_p^2}{2\mu_s \nu_p E}\rho g R z + u_0$ and $u_r(r,z) = u_\theta(r,z) = 0$. Thus, we obtain a Janssen's like redirection phenomenon due to a Poisson's ratio effect with a local Janssen's parameter $K_{el} = \sigma_{rr}(r,z)/\sigma_{zz}(r,z)$, being for large depths:

$$K_{el} = \nu_p/(1-\nu_p) \tag{2}$$

Thus, in the context of isotropic elasticity, the dependence of $K$ on packing fraction can be interpreted as an increase of the Poisson's ratio $\nu_p$ with packing fraction. Note a close derivation of stress distribution in an elastic column for $\lambda \gg R$ (this condition is not realized in our experiment) obtained by Evesque and de Gennes [14].

In order to get the whole stress saturation curve, finite element numerical simulations [15] were performed. The column is modelled as an isotropic elastic medium. We vary the friction at the walls $\mu_s$, the Young modulus $E$ and the Poisson coefficient $\nu_p$. We imposed a rigid, either perfectly stick or perfectly slip bottom. We found no appreciable difference between these two previous cases. The condition $\sigma_{rz} = \mu_s \sigma_{rr}$ is imposed everywhere at the walls. The cylinder is modelled as a steel elastic medium. As long as the Young modulus $E$ of the elastic medium is less than 500 MPa, which is usually the case for granular media, we found no dependence of the results on $E$. We verified that in all the simulations performed, there is no traction in the elastic medium, so that this can actually be a model for a granular material. We observed indeed that for low friction coefficients i.e. $\mu_s < 0.5$, the stress curve at the bottom of the column could not be separated from Janssen's universal curve. Note that the presence of a bottom imply that the effective Janssen's parameter $K_{el}^{eff}$ extracted from Janssen's scaling is higher than the $K_{el}$. We have in fact : $K_{el}^{eff} = aK_{el} + b$ with $a = 1.029 \pm 0.002$ and $b = -0.008 \pm 0.001$. If we use the correspondence with the effective Janssen's constants obtained experimentally this would yield an empirical relation between Poisson ratio $\nu_p$ and packing fraction $\nu$ as: $\nu_p \simeq 2.3(\nu - 0.41)$ with a precision of 5%. Note that the largest packing fraction $\nu = 0.645 \pm 0.005$ would give a Poisson ratio $\nu_p = 0.54 \pm 0.03$ marginally larger than the limit value of 1/2. Note that i) nothing really insures that our piling preparation is strictly isotropic and ii) in spite of the careful procedure, we are never absolutely sure that all the friction forces at the wall are actually fully mobilized upwards. We also performed simulations with an overweight (which is taken perfectly stick or slip with no appreciable difference), using brass mechanical parameters. We found that the resulting curve is very similar to the experimental one, but the experimentally observed maxima are consistently 30 *to* 40 *times larger* (see Fig. 3b). This is what we call the "*giant overshoot effect*"! We find from numerical computations the same qualitative phenomenology as in the experiment i.e. the computed values of the overshoot $M_{max}$ rescaled by the saturation mass $M_{sat}$, increase both with the friction at the walls and with the Poisson coefficient (i.e. with the effective Janssen's parameter). Basically, we recover in the elastic case, the same paradox as the experimental situation. We now understand it as a consequence of the boundary condition imposed experimentally by the overweight i.e. an almost constant displacement on the surface : $u_z(r) = u_0$. When $\nu_p$ or $\mu_s$ is decreased to zero (i.e when $M_{sat}$ increases), the parabolic part of the asymptotic vertical displacement becomes negligible and therefore, the displacement on the surface is close to the asymptotic value. This leads to a flat pressure profile.

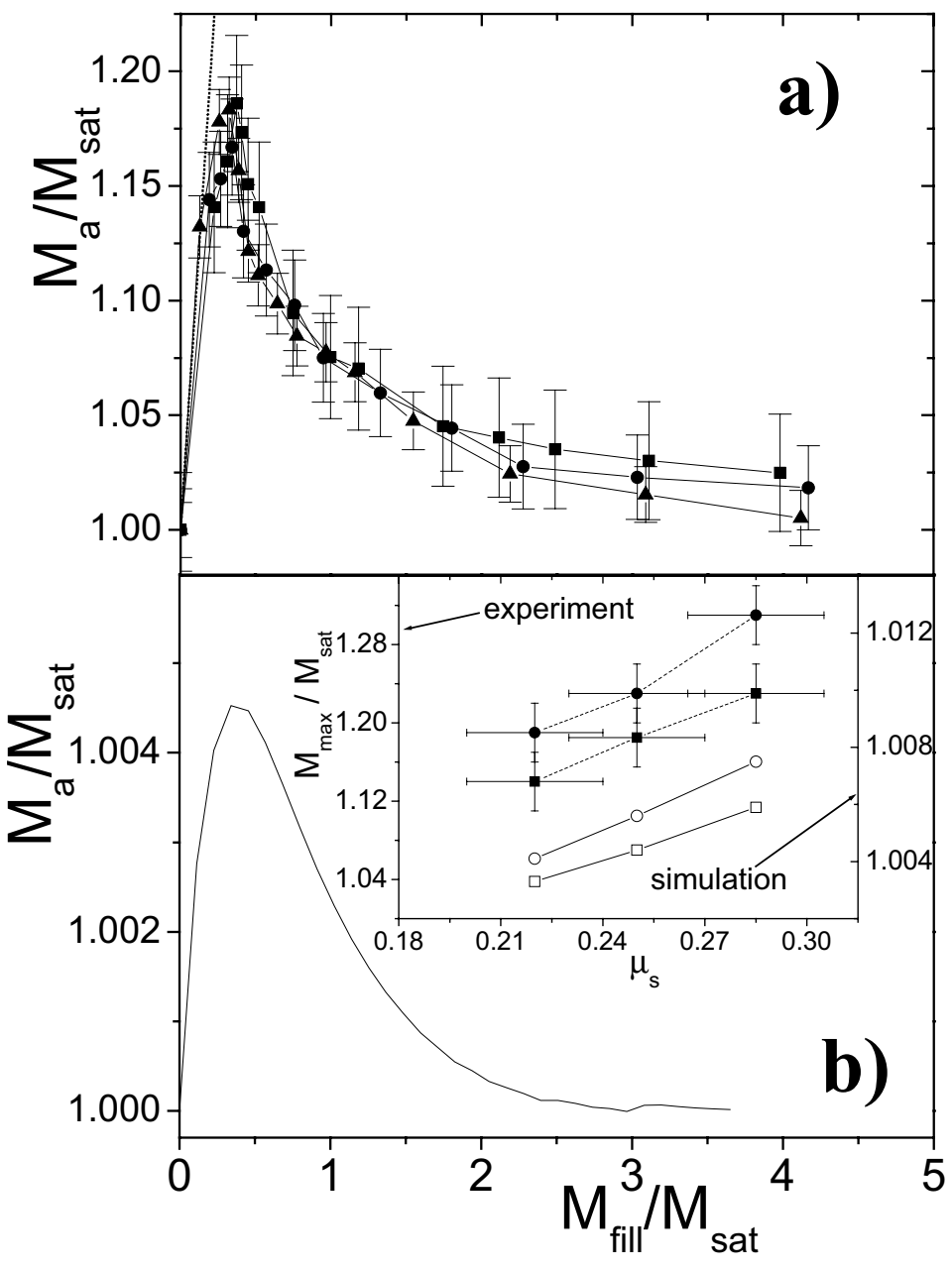

FIG. 3: a: Apparent mass $M_a$ vs. filling mass $M_{fill}$, rescaled by the saturation mass $M_{sat}$, for loose packings in medium-rough columns of 3 diameters (38 mm (squares), 56 mm (circles), 80 mm (triangles)) with an overweight equal to $M_{sat}$; the dotted line is the hydrostatic curve. b: Simulation of the experiment of Fig. 3a, for an elastic medium characterized by the same saturation mass (Poisson coefficient $\nu_p = 0.46$) and the same friction at the walls ($\mu_s = 0.25$). Inset: maximum mass $M_{max}$ rescaled by saturation mass $M_{sat}$ vs. static coefficient of friction at the walls $\mu_s$, in experiments made on loose (squares) and dense (circles) packings in 38 mm diameter columns, and in simulations for elastic media of Poisson coefficients $\nu_p = 0.45$ (open squares) and $\nu_p = 0.49$ (open circles); the left vertical scale is used for the experimental data, the right vertical scale is used for the simulation data.

Understanding the amplitude of this overshoot effect is still an open question but we propose here a qualitative vision based on stress induced anisotropy. Small mechanical restructuring due to displacements close to the overweight at the top could be capable to produce local anisotropy in a way that the vertical direction would becomes stiffer due to the preferential increase in the number of granular contacts. A consequence of this anisotropy is that the overshoot amplitude can be increased. This is what we verified numerically using a model of orthotropic elasticity [15] with the stiff axis taken in the vertical direction. But importantly, we need also to assume that this structuring effect has a finite spatial extend since for large column heights the original saturation limit is recovered.

In summary, we performed series of precise experiments on a granular material confined in cylindrical columns. We obtained for the first time the universal scaling predicted by Janssen in 1895 for the stress saturation curve when the filling mass is increased. We show that the screening parameter (the effective Janssen's constant) is an increasing function of the average packing fraction. By comparison with isotropic elastic theory we provide a one to one correspondence between the Poisson's ratio and granular packing fraction. We study systematically the overshoot effect, i.e. when a mass equal to the saturation mass is added to the top of the packing we have a non-monotonic variation of the apparent mass as a function of the filling mass. This effect is reproduced qualitatively by isotropic elasticity but of a scale 30 to 40 times smaller. We suggest a qualitative interpretation of this effect based on stress induced anisotropy. Therefore along these lines, this giant overshoot effect can be seen as a strong test for any theoretical model describing the statics of granular assemblies under various loading histories and boundary conditions.

We acknowledge discussions with P. Claudin and J. Socholar, and thank J. Lanuza for technical assistance.

[1] *Physics of Dry Granular Media*, ed. by H.J. Herrmann, J.-P. Hovi and S. Luding, Kluwer Acad. Publisher (1998).

[2] M.Oda, Mechanics of Materials **16**, 35 (1993).

[3] F. Radjai, D.E. Wolf, M. Jean and J.J. Moreau, Phys. Rev. Lett. **80**, 61 (1998).

[4] A.J. Liu, S.R. Nagel, Nature **396**, 21 (1998).

[5] M.E. Cates, J.P. Wittmer, J.-P. Bouchaud and P. Claudin, Phys. Rev. Lett. **81**, 1841 (1998).

[6] D.M. Wood, *Soil Behaviour and Critical State Soil Mechanics* (Cambridge University, Cambridge, England, 1990).

[7] L. Vanel et al., Phys. Rev. E **60**, R5040 (1999).

[8] J.P. Wittmer, M.E. Cates, P. Claudin, J. Phys. I **7**, 39 (1997); J.P. Wittmer et al., Nature **382** , 336 (1996).

[9] G. Reydellet, E. Clément, Phys. Rev. Lett. **86**, 3308 (2001); J. Geng et al., Phys. Rev. Lett. **87**, 035506 (2001); D. Serero et al., Eur. Phys. J. E, **6**, 169 (2001).

[10] L. Vanel, E. Clément, Eur. Phys. J. B **11**, 525 (1999). L. Vanel et al., Phys. Rev. Lett. **84**, 1439 (2000).

[11] H.A. Jannsen, Zeitschr. D. Vereines Deutscher Inginieure **39**, 1045 (1895).

[12] R. Peralta-Fabi, C. Malaga, R. Rechtman, Europhys. Lett, **45**, 76 (1999).

[13] S. N. Coppersmith et al., Phys. Rev. E **53**, 4673 (1996).

[14] P. Evesque, P.-G. de Gennes, C. R. Acad. Sci. Paris, **326** IIb, 761 (1998).

[15] with CAST3M, http://www.castem.org:8001